\documentclass{bcp_arch}
\usepackage{graphics}
\usepackage{graphicx}
\usepackage{amsmath}

\def\LaTeX{L\kern -.36em\raise .3ex\hbox{\sc a}\kern -.15em T\kern -.1667em%
\lower .7ex\hbox{E}\kern -.125em X}

\font\tt=cmtt10
\begin{document}

\mathclass{Primary 92D25; Secondary 82C26 and 82C31.}
\thanks{This research was in part supported by the US NSF grant DMR-0308548 
and (for M.M.) by Swiss NSF fellowship 81EL-68473.
M.M. gratefully acknowledges the support of the Alexander von Humboldt Foundation 
(fellowship IV-SCZ/1119205 STP). }
\abbrevauthors{M. Mobilia et al.}
\abbrevtitle{Stochastic predator-prey models}

\title{Spatial stochastic predator-prey models}

\author{Mauro Mobilia, $^{1,2}$ Ivan T. Georgiev,$^{2}$ and Uwe C. T\"auber $^{2}$}
\address{$^{1}$ Arnold Sommerfeld Center for Theoretical Physics and Center for NanoScience\\
 Department of Physics, Ludwig-Maximilians-Universit\"at M\"unchen, D-80333 Munich, Germany\\
$^{2}$ Department of Physics and Center for Stochastic Processes in Science and Engineering \\
Virginia Polytechnic Institute and State University, Blacksburg, Virginia 24061-0435, USA\\ 
{\tt E-mail:mauro.mobilia@physik.lmu.de,ivantgeorgiev@gmail.com,tauber@vt.edu}
}
\maketitlebcp

\abstract{We consider a broad class of stochastic lattice predator-prey models, whose main
features are overviewed. In particular, this article aims at drawing a picture
of the influence of spatial fluctuations, which are not accounted for by the deterministic rate equations, on the properties of the stochastic models. Here, we outline
the robust scenario obeyed by most of the lattice predator-prey models with an interaction
{\it \`a la Lotka-Volterra}. We also show how a drastically different behavior can emerge as the 
result of a subtle interplay between long-range interactions and a nearest-neighbor exchange
process. 
}
\section*{1. Introduction.}

Understanding biodiversity is a central challenge in modern evolutionary and theoretical
biology [May]. In this realm, studying population dynamic models to understand and identify the mechanisms allowing for coevolution of competing species is an important topic. The latter is classically addressed
by considering deterministic nonlinear differential equations. Within this approach, the set of equations 
devised many decades ago by Lotka and Volterra [Lot] is certainly a paradigm. These authors considered two coupled nonlinear differential equations mimicking the evolution of a two-species competing system. Hence, within their model, Lotka and Volterra demonstrated that the coexistence of the species always occurs and that the densities of the populations regularly oscillate in time. In spite of their popularity, the Lotka-Volterra equations
have often been criticized as being biologically unrealistic and mathematically unstable [May].
Actually, to gain some more realistic and fundamental understanding
on population coevolution and biodiversity, it is important to take into account discrete spatial degrees of freedom by going beyond the deterministic picture. In this context, various  stochastic predator-prey models have recently been investigated.
In this article, based on Refs.~[Mob1,Mob2] where more details can be found, we aim at drawing a brief overview of the robust properties of these lattice systems with Lotka-Volterra interactions. Also, by considering a simple ecological model, we will show how the subtle interplay between the degree of mixing of a stochastic system and long-range interaction can give rise to surprising features.

\section*{2. Deterministic rate equations.} It is useful to first review some results of the deterministic approach for the two-species system, where predators $(A)$ and prey $(B)$
interact according to the following reactions: $A\to \oslash$ (death rate $\mu$), 
$A+B\to A+A$ (predation rate $\lambda$) and $A\to A+A$ (reproduction rate $\sigma$). Neglecting
any spatial variations of the densities $a(t)$ and $b(t)$ of predators and prey, respectively, one 
obtains the classical Lotka-Volterra equations:
\begin{eqnarray}
\label{LV}
\dot{a}=\lambda a(t)b(t)-\mu a(t); \quad \dot{b}=\sigma b(t)-\lambda a(t)b(t). 
\end{eqnarray}
A linear stability analysis of these equations show that the densities oscillate around the
center (neutrally stable) fixed point $(a_c,b_c)=(\sigma/\lambda,\mu/\lambda)$ with a characteristic frequency $f=\sqrt{\mu \sigma}/2\pi$. In addition, the existence of a conserved first integral
$K(t)=\lambda[a(t) + b(t)]-\sigma \ln{a(t)}-\mu \ln{b(t)}$ implies oscillatory kinetics and coexistence in the whole phase portrait, which is characterized by cyclic trajectories.

As the above deterministic cycles are unstable against any perturbations and the solutions of
Eqs.~(\ref{LV}) display amplitudes depending on the initial conditions, which is clearly an unrealistic 
feature, the rate equations are often rendered more realistic by including growth-limiting terms.
For instance, by assuming that the prey carrying capacity is $\rho$, one is led to the following equation for species $B$:
\begin{eqnarray}
\label{LVcar}
\dot{b}=\sigma b(t)[1-b(t)/\rho]-\lambda a(t)b(t). 
\end{eqnarray}
In this case, the new rate equations exhibit an extinction threshold
and have two stable fixed points: (i) $(0,\rho)$, corresponding
to a system full of prey and extinction of the predators, which is a stable node when $\lambda<\lambda_c=\mu/\rho$ (extinction threshold); and (ii)
$([1-\mu/\lambda\rho]\sigma/\lambda, \mu/\lambda)$, which is associated with the coexistence phase and is either a stable node (near the extinction threshold) or a focus (deep in the coexistence phase). 
The existence of a Lyapunov function ensures that these fixed points are actually {\it globally stable} [May,Mob1,Mob2].

\section*{3. The stochastic lattice Lotka-Volterra model (and its variants).}
\begin{figure}
\begin{center}
\includegraphics[scale=0.57]{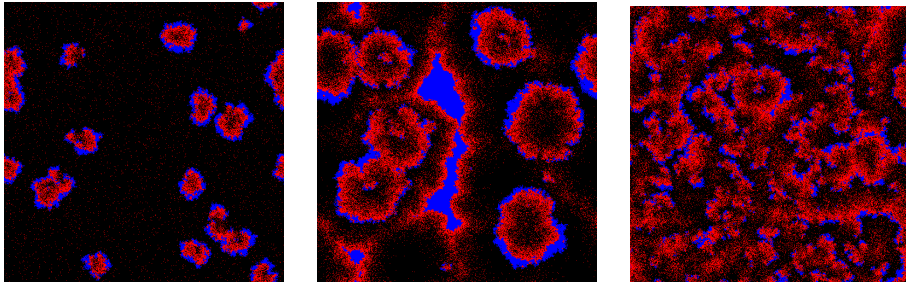}
\caption{(color online) Reproduced from [Mob2]. Snapshots of the time evolution (time increases from left to right)
of the SLLVM on a $512\times 512$ lattice, with rates $\sigma=4.0, \mu=0.1, \lambda=2.2$.
The red, blue and dark dots respectively represent the prey, predators and empty sites.
Initially the system is homogeneous with densities $a(0)=b(0)=1/3$. 
\label{snap}}
\end{center}
\end{figure}

To probe the deterministic approach and gain some understanding on the role of spatial
fluctuations and correlations,  we have studied a stochastic lattice Lotka-Volterra model (SLLVM). The latter is formulated in the natural language of reaction-diffusion systems and is defined on a $d-$dimensional lattice, on which the above Lotka-Volterra reactions are implemented. In our modeling, we mimic
spatial limitation of the resources by assuming that each lattice site can be at most occupied by one predator
or one prey (site restriction). 
Monte Carlo simulations of the SLLVM show that, in dimensions $d>1$, the phase portrait displays {\it qualitatively} the behavior predicted by the rate equations with growth-limiting term (see Fig.1 of [Mob2]). 
In fact, it is found that above some critical threshold both predators and prey coexist. The related fixed point is either a node or, for high predation rate, a focus which is associated with spiraling flows. In the latter case, as shown in Fig.~(\ref{snap}), rich spatiotemporal patterns of persistent predator-prey ``pursuit and evasion'' waves (see e.g. Murray's book [May]) develop and translate into erratic damped population density oscillations [Mob2]. Obviously these features are not captured by the rate equations (\ref{LV},\ref{LVcar}).
In {\it finite} systems, the quasiperiodic fluctuations appear on a global scale, with amplitudes vanishing with the system size. We have analyzed the spatial structure and the time evolution of the above complex patterns by computing the stationary two-point correlation functions and the power spectrum [Mob2]. This has led us to an understanding of the typical size of the clusters reported on Fig.~(\ref{snap}, rightmost) and has shown that the characteristic frequency of the SLLVM is markedly smaller than the ones predicted by the deterministic equations [Mob2].
A completely different picture emerges when the reactive fixed point is a node, just above the 
extinction threshold ($\lambda_c$) for the predators. Here, one observes clouds of predators effectively diffusing in a sea of prey (see Fig.~3 of [Mob2]). At the critical value, the system undergoes an {\it active-to-absorbing 
 phase transition}. Various critical exponents were computed  (both in 2D and 3D)
and found to be in agreement with those of the {\it directed percolation} (DP) [Jans].
By deriving an equivalent field-theory action 
from the master equation of the SLLVM and by mapping it (near $\lambda_c$) onto Reggeon field-theory,
we provided analytic arguments supporting the numerical indications
that such a transition is in the DP universality class [Mob2]. Similar DP exponents were found numerically in many other stochastic predator-prey models (see e.g.~[Ant]).
As a further probe of the robustness of the above scenario, we have 
considered different variants of the stochastic model, namely the SLVVM supplemented with 
(i) diffusing predators and prey (with same diffusivities); (ii) biased diffusion of both species; (iii) predators ``following" the prey, with biased hoping rates towards neighboring sites occupied by prey. In all these cases, the above features were reproduced.
Let us also note that for SLLVM without site restriction it was shown that predators and prey always coexist in 1D and 2D [Wash].
\section*{4. Mean-field behavior through short-range stirring.} 
We have outlined the robust properties of most lattice predator-prey models. Among these characteristics,
we have seen that deep in the coexistence phase the systems display complex clusters of activities at the interfaces of which the dynamics takes place. These features are not affected by the diffusion of either predators or prey. On the other hand, 
it is important to identify if there are ecological mechanisms able to render stochastic predator-prey models more tractable by means of deterministic equations. This requires to find  processes allowing to efficiently
stir the system by bringing the reactants within the interfaces of the clusters.
Arguably, the simplest candidate is the process  
allowing any neighboring sites to exchange their content (a prey can avoid an incoming predator): 
$X+Y \to Y+X$, with $X,Y \in \{A,B, \oslash\}$ and $X\neq Y$. To test the efficiency
of this ingredient, we have focused on a model whose mean-field (MF) behavior is markedly different from that of Eqs.~(\ref{LV},\ref{LVcar}). Actually, it is natural to
 split the Lotka-Volterra predation reaction by introducing two independent time scales and the new processes:
(i) reproduction of the predators in the vicinity of a prey ({\it favorable environment}), according to the {\it next-nearest-neighbor} (NNN) reaction $A+\oslash + B \to A+A+B$ (with rate proportional to $\delta$); and (ii) consumption of a prey by a neighboring predator, as $A + B \to \oslash+B$ (with rate proportional to $\eta$). The MF equations for this model, which can be viewed as comprising also a nearest-neighbor reproduction reaction ($A+\oslash  \to A+A$) taking place on a much longer time scale,
read [Mob1]
\begin{eqnarray}
\label{LV_NNN}
\dot{a}=\delta a(t)b(t)[1-a(t)-b(t)]-\mu a(t); \quad \dot{b}=\sigma b(t)[1-a(t)-b(t)]-\eta a(t)b(t). 
\end{eqnarray}
In stark contrast to Eqs.~(\ref{LV},\ref{LVcar}), these  
equations admit {\it two stable fixed points} above a given threshold $\delta_c$: 
one is (always) a node and corresponds to an absorbing steady state
(system full of prey), while the other fixed point (either a node or a focus) is
 reactive (coexistence of prey, with a density $<1/2$, and predators). Hence, according to MF theory, for $\delta>\delta_c$
this model undergoes a {\it first-order phase transition}.
To check when these MF predictions actually apply to the stochastic lattice model,
the latter has been numerically simulated  in the presence of the 
exchange process (with stirring rate proportional to ${\cal D}$) [Mob1].
\begin{figure}
\begin{center}
\includegraphics[scale=0.6]{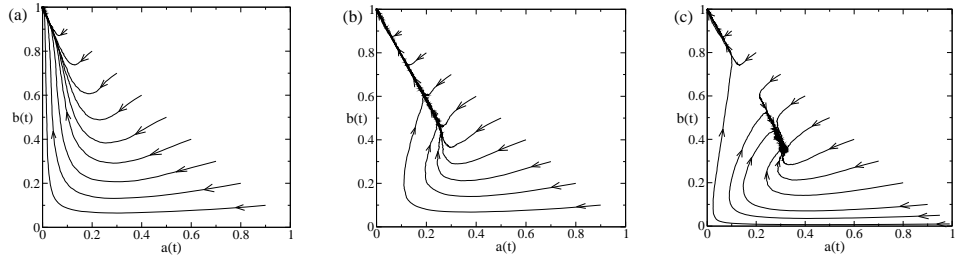}
\caption{(color online) Reproduced from [Mob1]. Effect of stirring on the phase portrait of 
the 2D stochastic lattice predator-prey with NNN interaction (rates $\eta = \mu = \sigma = 1$,
$\delta = 10$, system size: $256 \times 256$).
>From left to right: `stirring' rate ${\cal D} = 0, 2, 5$.
\label{stir}}
\end{center}
\end{figure}
It turns out that there is  a subtle interplay
between the NNN interaction and the exchange process.
As illustrated in Fig.~\ref{stir}, the latter can be summarized as follows:
(a) For vanishing mixing (${\cal D}$ small compared to the other rates), 
fluctuations have a drastic effect and invalidate the MF picture
 in dimensions $1 < d \leq 4$, where the system undergoes an active-to-absorbing state
transition belonging again to the DP universality class.
(b) When one allows for short-range particle exchange (${\cal D} > 0$),
the phase portrait flows change dramatically (Fig.~\ref{stir}, center).
(c) When the exchange process becomes sufficiently fast (e.g.
${\cal D} \approx \delta$) a reactive fixed point is also available (in any dimension), as demonstrated in
Fig.~\ref{stir} (right). In this case, the system undergoes a first-order phase transition
as predicted by the MF theory and, for `fast' stirring, the latter becomes
very accurate. We have also found that the stable reactive fixed point
is either a node or a focus. In the latter case, the coexistence phase 
is again characterized by  moving activity fronts but, as the system is 
more mixed, these clusters appear less prominent than in Fig.~\ref{snap} [Mob1].

\section*{5. Conclusion.} 
We have outlined the robust features of a class of stochastic lattice
models with Lotka-Volterra interactions and discussed how their deterministic 
(mean-field) descriptions are altered by spatial fluctuations and correlations. We have also shown that the rate equations can aptly describe the dynamics of a stochastic model in the presence of an efficient short-range exchange process. We have illustrated this point by considering 
a system with NNN interaction which exhibits either a first- or second-order phase transition, 
depending on the stirring rate.

\references{Ant}
{
\item{[Ant]} T. Antal and M. Droz, {\it Phase transitions and oscillations in a lattice prey-predator model}, Phys. Rev. E {\bf 63} (2001), 056119; 
M. Kowalik, A. Lipowski and A. L. Ferreira, {\it Oscillations and dynamics in a two-dimensional prey-predator system}, {\it ibid.} {\bf 66} (2002), 066107.
\item{[Jans]} H. K. Janssen, {\it On the non-equilibrium phase-transition in reaction-diffusion
systems with an absorbing stationary state}, Z. Phys. B: Condens. Matt {\bf 42} (1981), 151;
H. Hinrichsen, {\it Non-equilibrium critical phenomena and phase transitions into absorbing 
states}, Adv. Phys. {\bf 49} (2000), 815.
\item{[Lot]} A. J. Lotka, {\it Undamped oscillations derived from the law of mass action}, J. Am. Chem. Soc. {\bf 42} (1920), 1595; V. Volterra, {\it Le\c cons sur la th\'eorie math\'ematique de la lutte pour la vie}, Gautiers-Villars, Paris, 1931.
\item{[May]}
R. M. May, {\it Stability and Complexity in Model Ecosystems},
  Princeton University Press, Princeton, 1973;
   J. Maynard Smith, {\it Models in Ecology}, 
  Cambridge University Press, Cambridge, 1974;
  J. D. Murray, {\it Mathematical Biology} Vols.~I/II, Third Edition, 
  Springer-Verlag, New York, 2002;
 D. Neal, {\it Introduction to Population Biology}, 
Cambridge University Press, Cambridge, 2004.
 \item{[Mob1]} M. Mobilia, I. T. Georgiev and U. C. T\"auber, {\it Fluctuations and correlations in lattice models for predator-prey interaction}, Phys. Rev. E {\bf 73} (2006), 040903(R).
  \item{[Mob2]} M. Mobilia, I. T. Georgiev and U. C. T\"auber, {\it Phase Transitions ans Spatio-Temporal
  Fluctuations in Stochastic Lattice Lotka-Volterra Models}, J. Stat. Phys, 
published online (DOI: 10.1007/s10955-006-9146-3). {\tt E-print:q-bio.PE/0512039}.
\item{[Wash]} M. J. Washenberger, M. Mobilia and U. C. T\"auber, {\it Influence of local
carrying capacity restrictions on stochastic predator-prey models}, 
J. Phys. Condens. Matter {\bf 18} (2006), XXX.
{\tt E-print:cond-mat/0606809.}
}

\end{document}